\documentclass[12pt]{article}
\usepackage{pic02}
\usepackage{hyperref}
\usepackage{url}
\usepackage{epsfig}
\usepackage{floatflt}
\def\be{\begin{equation}}
\def\ee{\end{equation}}
\def\bea{\begin{eqnarray}}
\def\eea{\end{eqnarray}}
\def\noi{\noindent}
\def\ds{\displaystyle}
\def\br{{\bf r}}
\def\etal{\mbox{\it et al.\,}}
\def\Journal#1#2#3#4{{#1} {\bf #2}, {#3} ({#4})}
\def\IJMPA{{\it Intl. J. Mod. Phys.} A}

\def\PLB{{\it Phys. Lett.}  B}

\def\PRD{{\it Phys. Rev.} D}

\def\EPJC{{\it Eur. Phys. J.} C}

\begin{document}

\title{\bf LOW $x$ PHYSICS AT HERA}

\author{Robin Devenish \\ \\
Representing the H1 \& ZEUS Collaborations. \\  \\ 
{\em Oxford University, Physics Department} \\
{\em Denys Wilkinson Bldg, Keble Rd}  \\
{\em Oxford OX1 3RH, UK}}
\maketitle

%
% photograph of author
%  This is where we will insert a photograph. To see what it would look like,
%  uncomment the following lines.
%
%\begin{figure}[h]
%\begin{center}
%
% include photograph for proceeding version
%
%\includegraphics[height=4.5cm]{einstein.eps}
%
% insert a fixed vertical spacing instead for the ArXiv preprint
%
\vspace{4.5cm}
%
%\end{center}
%\end{figure}

\baselineskip=14.5pt
\begin{abstract}
Data on low $x$ physics from ZEUS \& H1 are presented and their interpretation
discussed. The focus is on the increasing hardness of the energy dependence
of inclusive $\gamma^*p$ scattering and certain diffractive processes as the
transverse size of the probe decreases.
\end{abstract}
\newpage

\baselineskip=17pt

\section{Introduction}

HERA remains a unique facility, the only $e^\pm p$ collider.
With $27.5\,$GeV $e^\pm$ beams on $920\,$GeV protons the centre of mass energy
is $318\,$GeV which also gives the upper bound on momentum transfer and thus 
the scale to which proton structure can be probed. Apart from extending the
study of deep inelastic scattering and partonic physics, HERA also allows
$\gamma^*p$ interactions to be study over a wide range of photon
virtuality down to almost real photons. It provides a
laboratory in which the high energy behaviour of cross-sections
may be studied as a function of the transverse size of the projectile.
This is the realm of `low $x$' physics. This brief survey covers: data
from H1 and ZEUS on inclusive and diffractive processes: their 
analysis using perturbative QCD and other models, particularly
colour diople models; the evidence or otherwise for universality and gluon
saturation at low $x$.

\section{Formalism and phase space}

\begin{figure}[htb]
\begin{center}
\begin{tabular}[t]{ll}
\epsfig{file=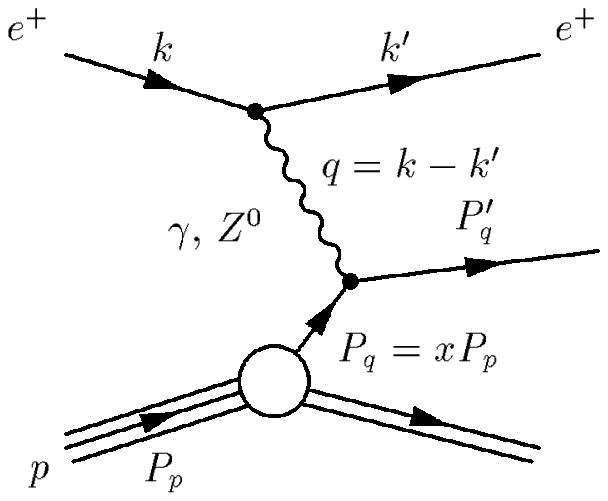,width=.3\textwidth,clip=} &
\epsfig{file=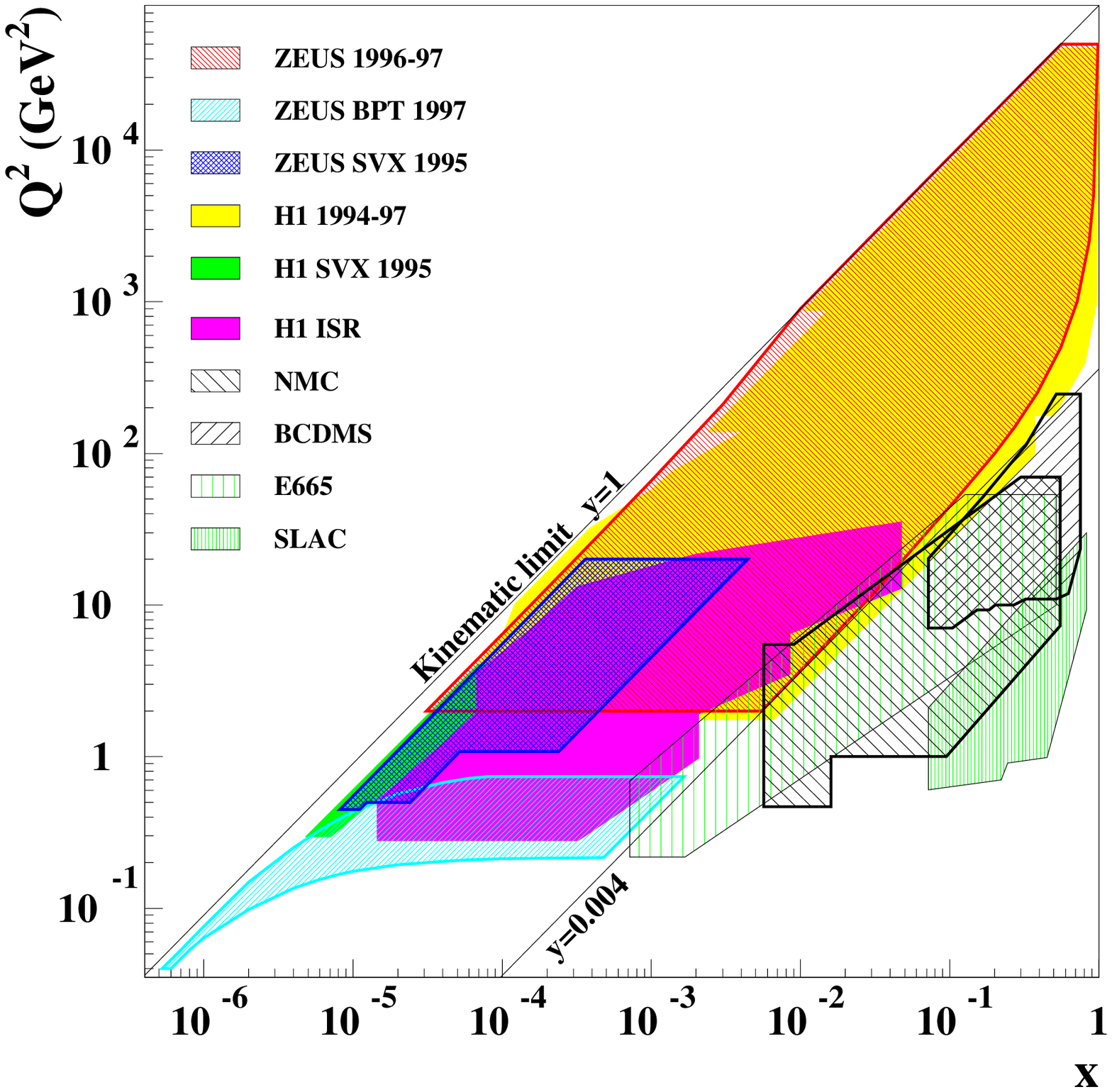,width=.6\textwidth} 
\end{tabular}
 \caption{\it LH:  Inclusive ep scattering and momenta. RH:
     Kinematic Plane for inclusive ep scattering up to HERA energies.
    \label{fig:kinreg} }
\end{center}
\end{figure}

\noi
Inclusive electron--proton scattering at HERA $ep\to eX$ is shown
in Fig.~\ref{fig:kinreg} -- LH. At a fixed centre of mass energy, 
$\sqrt{s}$, where $s=(k+p)^2$)
the process is described by two kinematic variables, $Q^2$, the
four-momentum transfer squared and Bjorken $x$ where
\be
Q^2=-q^2=-(k-k^\prime)^2:~~~x=Q^2/(2p.q).
\ee
The inelasticity, $y=(p.q)/(p.k)$, the fractional energy loss of the 
electron in the proton rest frame, is related to $x$ and $Q^2$ by
$Q^2=sxy$. The primary measured quantity is the double differential
cross-section which, for $Q^2 << M^2_Z$, may be written in terms
of two structure functions
\be
{d^2\sigma\over dxdQ^2}={2\pi\alpha^2\over xQ^4}\left[
(1+(1-y)^2)F_2(x,Q^2)-y^2F_L(x,Q^2)\right].
\label{eqn:d2xq2}
\ee
The contribution from $y^2F_L$, the longitudinal structure function,
is small and will be ignored in most of what follows.
The inclusive scattering process may also be considered in terms
of the total cross-section for $\gamma^*p$ scattering.
At small $x$
\be
\sigma^{tot}_{\gamma^*p}(W^2,Q^2)={4\pi^2\alpha\over Q^2}F_2(x,Q^2),
~~~{\rm and}~~~W^2\approx Q^2/x,
\label{eqn:f2-sig}
\ee
where $W^2=(p+q)^2$ is the centre of mass energy squared of $\gamma^*p$
system. So the structure function at small $x$ gives 
$\ds \sigma^{tot}_{\gamma^*p}$ at high CM energies.

\noi
The $x,Q^2$ region in which inclusive measurements have been made is shown
in Fig.~\ref{fig:kinreg} -- RH. The strong correlation between $x$ and $Q^2$
follows from the constraint $Q^2=sxy$ and the kinematic limit at HERA
is given by the line $y=1$. Fixed target measurements are in the lower
right-hand quadrant of the plot. The region of interest for this talk is 
roughly $x<0.01$, and $Q^2<250\,$GeV$^2$. 

\section{Contexts}
\label{sec:pqcd}

The study of deep inelastic lepton scattering (DIS) and the development of QCD
have been intimately related and this continues in the low $x$ region
opened up by HERA.\footnote{Much more detail and references to original papers
may be found in review article of ref.~\cite{cdd}.} 
The success of pQCD for hard processes relies on 
factorisation theorems which, in the case of DIS, enables the 
structure function $F_2$ to be written in the form
\be
F_2(x,Q^2)=\sum_f e_f^2\,xq_f(x,Q^2),~~{\rm with}~~
{\partial q_i(x,Q^2)\over \partial\ln Q^2}\sim \sum_jq_j\otimes P_{ij},
\label{eqn:dglap}
\ee
where $q_f$ is the momentum density function for quark
of flavour $f$ and charge $e_f$, $P_{ij}$ are the QCD splitting functions and
the sums run over $q$ and $\bar{q}$. At low $x$ $F_2$ is largely
given by the flavour singlet $q\bar{q}$ sea contribution and the 
evolution of this term is coupled to that of the gluon. The behaviour
of both is dominated by the singular behaviour of the gluon splitting 
function $P_{gg}\sim 1/x$, giving rise to an increase in the gluon
density, the $q\bar{q}$ sea and hence $F_2$ at low $x$. 

\noi
The DGLAP equations (Eq.~\ref{eqn:dglap}), which have been 
calculated to next-to-leading order (NLO)
in $\alpha_S$ and partially to next-to-next-to-leading (NNLO), represent
a particular, physically motivated, choice of how higher order corrections
are summed. The most general behaviour of the splitting functions may be 
represented by double sum over terms of the form $(\ln Q^2)^n(\ln1/x)^m$
and there are other possible choices of summation that may be relevant at
low $x$. One of earliest results of pQCD as applied to DIS 
was derived by taking the leading log terms in both $Q^2$ and $1/x$, giving 
\be
F_2(x,Q^2) \sim \exp\sqrt{(12\alpha_s/\pi)\ln(1/x)\ln(Q^2/Q^2_0)}.
\label{eqn:dlla}
\ee   
Another approach, taken by Balitsky et al (BFKL), involves summing
the $\ln(1/x)$ terms and gives the striking prediction that 
$F_2\sim x^{-\lambda}$, where $\lambda\approx 0.5$ at leading order
and fixed $\alpha_S$. For running $\alpha_S$ and NLO, the corrections
to this result are large and are still subject to much debate. Despite the
uncertainty surrounding the exact prediction, the BFKL approach has been
of seminal importance in low $x$ QCD dynamics. The DGLAP and BFKL approaches  
represent two `extreme' choices in how  the transverse momenta of the 
radiated gluons are ordered. All calculations agree that 
gluon dynamics dominates at low $x$ and that the gluon density will increase
as $x$ decreases. At some point the gluon density will be so large that 
non-linear gluon recombination effects will need to be taken into account. 
A simple estimate suggests that such effects may occur for $Q^2$ values
for which the gluon-gluon cross-section times the gluon density is of order 
the proton size:
\be
{\alpha_S(Q^2)\over Q^2}xg(x,Q^2)\sim \pi R^2,
\ee
where $R$ is the proton radius and $xg$ the gluon density, thus slowing
the rise of $F_2$.

\noi
Hadronic total cross-sections involve soft physics which cannot be
calculated perturbatively. Regge theory provides a framework for describing
the high energy behaviour of total cross-sections. One
expects $\sigma^{tot}_{hadrons}\sim A\cdot W^{2(\alpha_P(0)-1)}$,
where $A$ is a process dependent constant and $\alpha_P(0)$, the intercept 
of the `Pomeron trajectory'\footnote{In principle $\alpha_P$ is determined
by the exchange of states with vacuum quantum numbers in 
the crossed channel.}, is process independent. Although the value of
$\alpha_P(0)$ is not given, the model has been successfully
applied to a wide range of hadron-hadron and real photoproduction data,
giving a fitted value of $1.093(2)$ \cite{DL1}.
If these ideas also apply to $\gamma^*p$ scattering, using 
Eq.~\ref{eqn:f2-sig}, one might expect
$F_2\sim x^{1-\alpha_P(0)}$ at small $x$. Using the hadronic value of
$\alpha_P(0)$ predicts a rather gradual rise $F_2\sim x^{-0.09}$, with
the exponent independent of $Q^2$. The BFKL calculation may be viewed
as an attempt to calculate $\alpha_P(0)$ perturbatively.
Quasi-elastic scattering processes, such as diffraction, should also
be dominated by the exchange of vacuum quantum numbers and the Pomeron, 
which is why they are important in unravelling the secrets of low $x$
dynamics.

\section{$F_2$ at low $x$}

\begin{figure}[htbp]
\begin{center}
\begin{tabular}[t]{ll}
\epsfig{figure=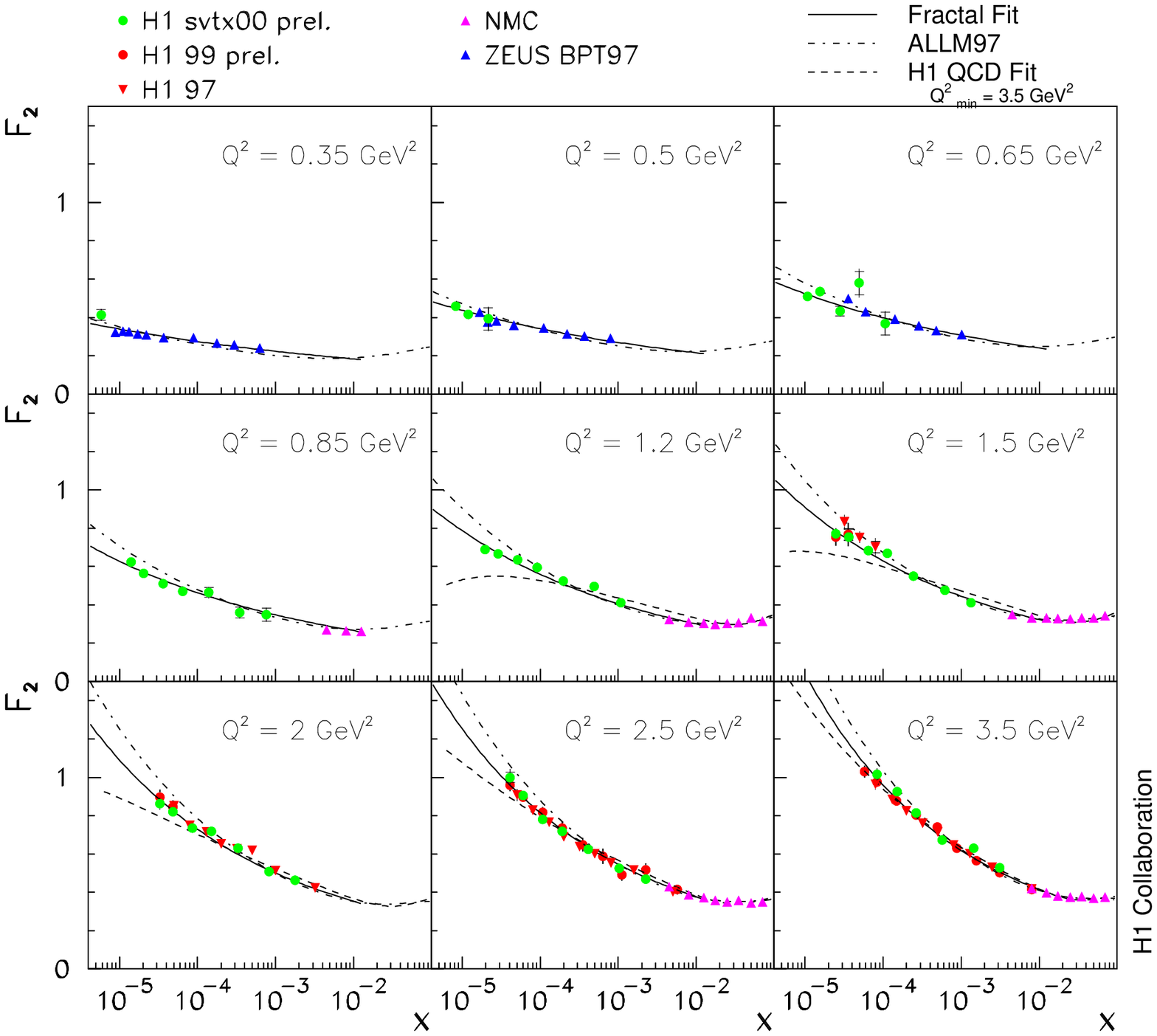,width=.55\textwidth} &
\epsfig{figure=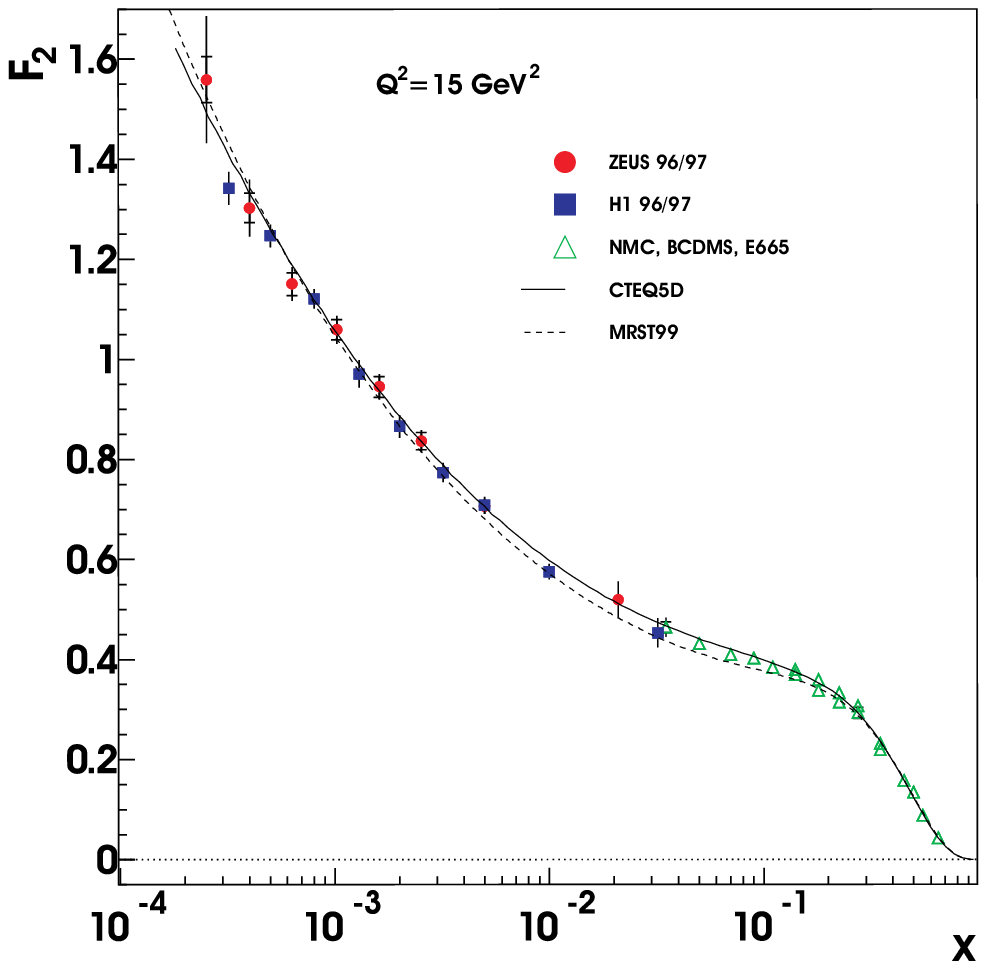,width=.35\textwidth} 
\end{tabular}
\caption{\it $F_2$ data from H1, ZEUS \& NMC as a function of $x$ in bins of 
$Q^2$.}
\label{fig:f2vsx}
\end{center}
\end{figure}

Fig.~\ref{fig:f2vsx} shows $F_2$ data from H1, ZEUS and the NMC
fixed target experiments, plotted as a function
of $x$ in bins of $Q^2$. The RH plot, for $Q^2=15\,$GeV$^2$ well 
into the deep inelastic range, shows $F_2$ rising steeply as $x$ decreases. 
The LH plots show
bins with $Q^2$ decreasing from $3.5$ to $0.35\,$GeV$^2$. Here the rise
is still evident at the larger $Q^2$ values, but diminishes rapidly as
$Q^2$ decreases below $\sim 1\,$GeV$^2$. From Eq.~\ref{eqn:f2-sig} the 
rapid rise of $F_2$ at small $x$ implies that
$\sigma^{tot}_{\gamma^*p}(W^2,Q^2)$ increases more rapidly with $W^2$
as $Q^2$ increases, which is not expected in the hadronic Regge framework.

\begin{figure}[htbp]
\begin{center}
\begin{tabular}[t]{ll}
\epsfig{figure=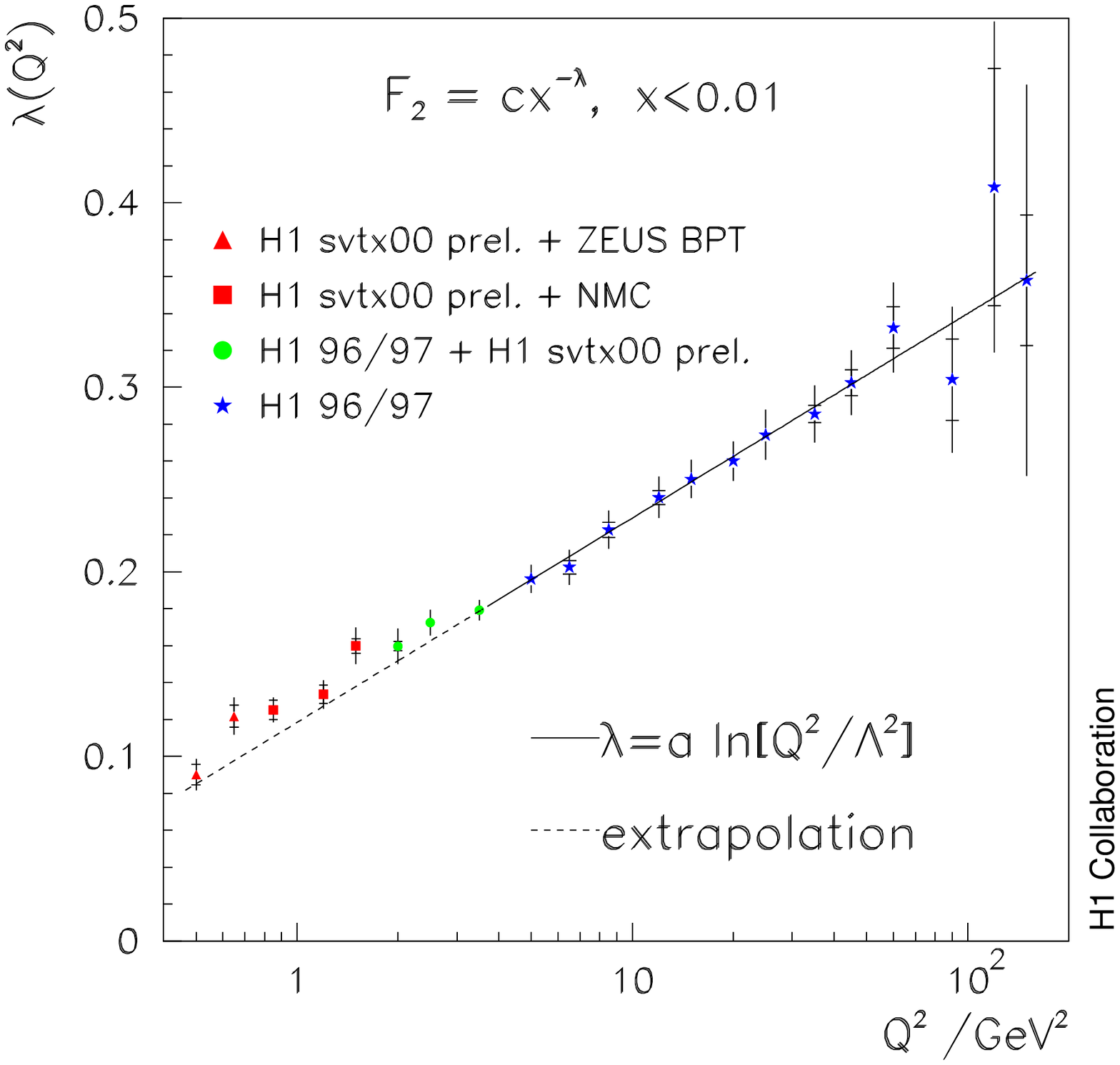,width=.4\textwidth} &
\epsfig{figure=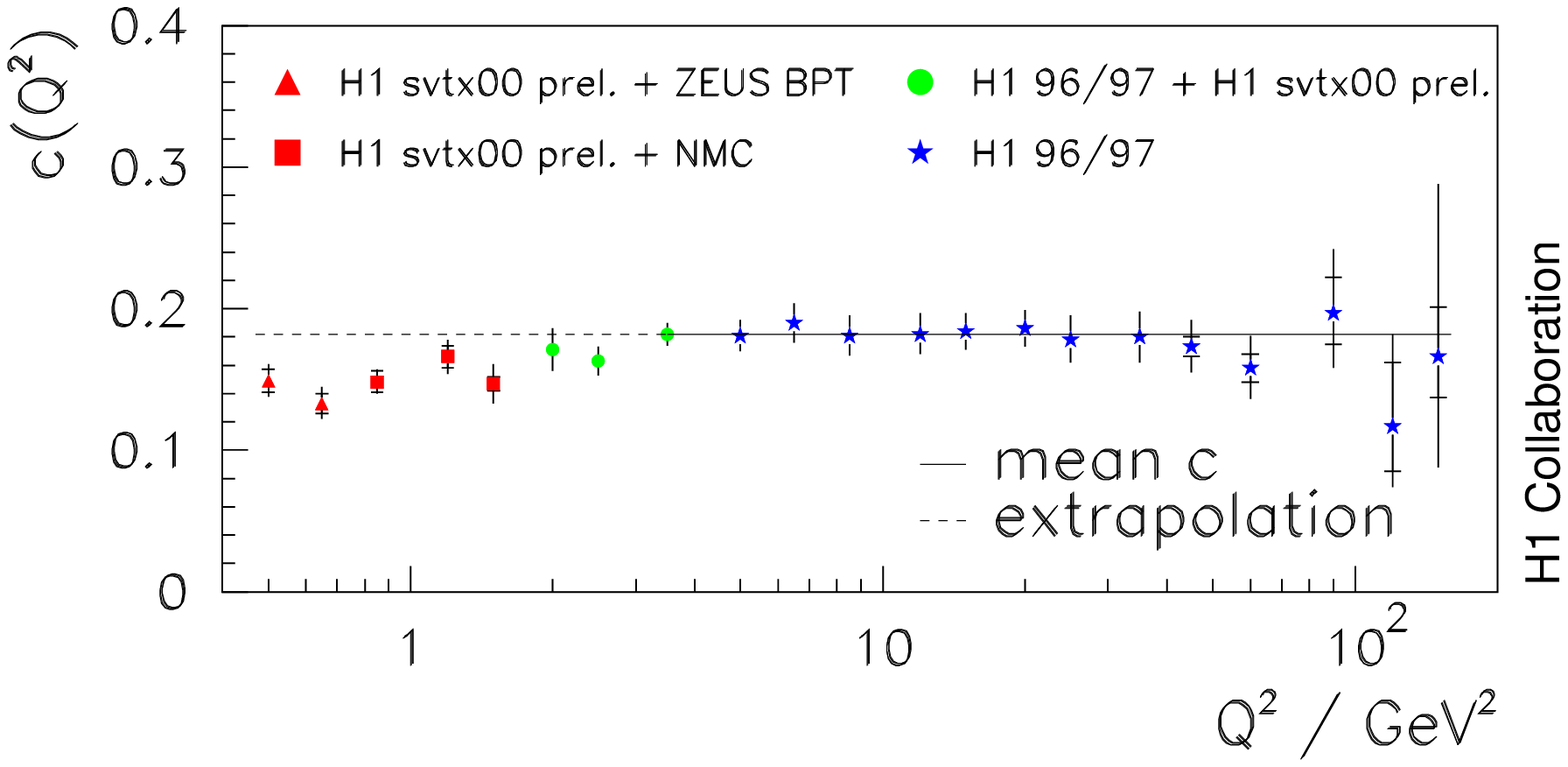,width=.4\textwidth} 
\end{tabular}
\caption{\it The logarithmic slope and constant versus $Q^2$ from the 
$F_2(x,Q^2)=c(Q^2)x^{-\lambda(Q^2)}$ fit to data with $x<0.01$. The $F_2$
data used is mostly from H1 with some very low $Q^2$ points from ZEUS.}
\label{fig:h1-lam}
\end{center}
\end{figure}

\noi
H1 have made a model independent study of the rise of $F_2$
\cite{h1-lambda}. Using data with $x<0.01$, the logarithmic slope
with respect to $\ln x$ is calculated using
\be
\lambda(Q^2)=-\left.{\partial\ln F_2\over \partial \ln x}\right|_{Q^2},
~~~~{\rm or}~~~F_2(x,Q^2)=c(Q^2)x^{-\lambda(Q^2)}.
\label{eqn:lam-def}
\ee 
The data from HERA are now sufficiently extensive and precise for 
$\lambda$ to be measured in bins of $Q^2$ over a range of $x$ values.
At fixed $Q^2$, $\lambda$ is found to be independent of $x$ for $x<0.01$.
Fig.~\ref{fig:h1-lam} shows $\lambda$ and $c$ as functions of $Q^2$ where
the errors have been calculated using the full systematic and
statistical errors of the $F_2$ data including correlations. At low values 
of $Q^2<1\,$GeV$^2$, the value of $\lambda$ is not far above the value 
expected from the hadronic Regge parameterisation, whereas at
larger values of $Q^2$, $\lambda$ rises almost linearly with $\ln Q^2$ to 
a value around $0.3$ at  $Q^2\approx100\,$GeV$^2$. 
The results for $c(Q^2)$ are roughly independent of $Q^2$ with a 
value of about $0.18$.  The data
show no indication that the rate of rise is starting to moderate at 
larger $Q^2$, as might be expected from gluon
saturation. Similar results for the logarithmic slope have been found in 
a preliminary analysis of ZEUS data~\cite{surrow}.  

\noi
Although the results from the $\lambda$ analysis support the hypothesis
of gluon dominance at low $x$, it is not possible from this study alone
to say which of the pQCD approaches fits best. More information
may be gained by applying the models discussed in Sec.~\ref{sec:pqcd} 
to the $F_2$ data over the full range of HERA and fixed target measurements. 
The most detailed fits are those
using NLO DGLAP to determine the nucleon parton momentum densities.
More details are given in the talk by Milstead~\cite{milstead}, but the
quality of the fits may be judged from Fig.~\ref{fig:f2vsx} which shows
that NLO DGLAP can describe $F_2$ for $Q^2$ values larger than 
$1.5-2\,$GeV$^2$ and all $x$ values. For other pQCD approaches, fits
to the low $x$ $F_2$ data of comparable quality are obtained
but the predictions for other observables, for example $F_L$,
differ quite considerably. Although $F_L$ has been extracted in the HERA
kinematic region \cite{h1-FL}, the data do not yet have the precision to
distinguish between the models. 

\section{Universality at low $x$?}

The $F_2$ data show gluon and $q\bar{q}$ sea dominance
at low $x$. As the influence of the valence quarks is also small, 
the possibility arises that this behaviour
could, in some sense, be `universal'. To test this idea the structure
function of another hadron must be extracted. Both ZEUS and H1 have
\begin{figure}[htbp]
\begin{center}
\begin{tabular}[t]{ll}
\epsfig{figure=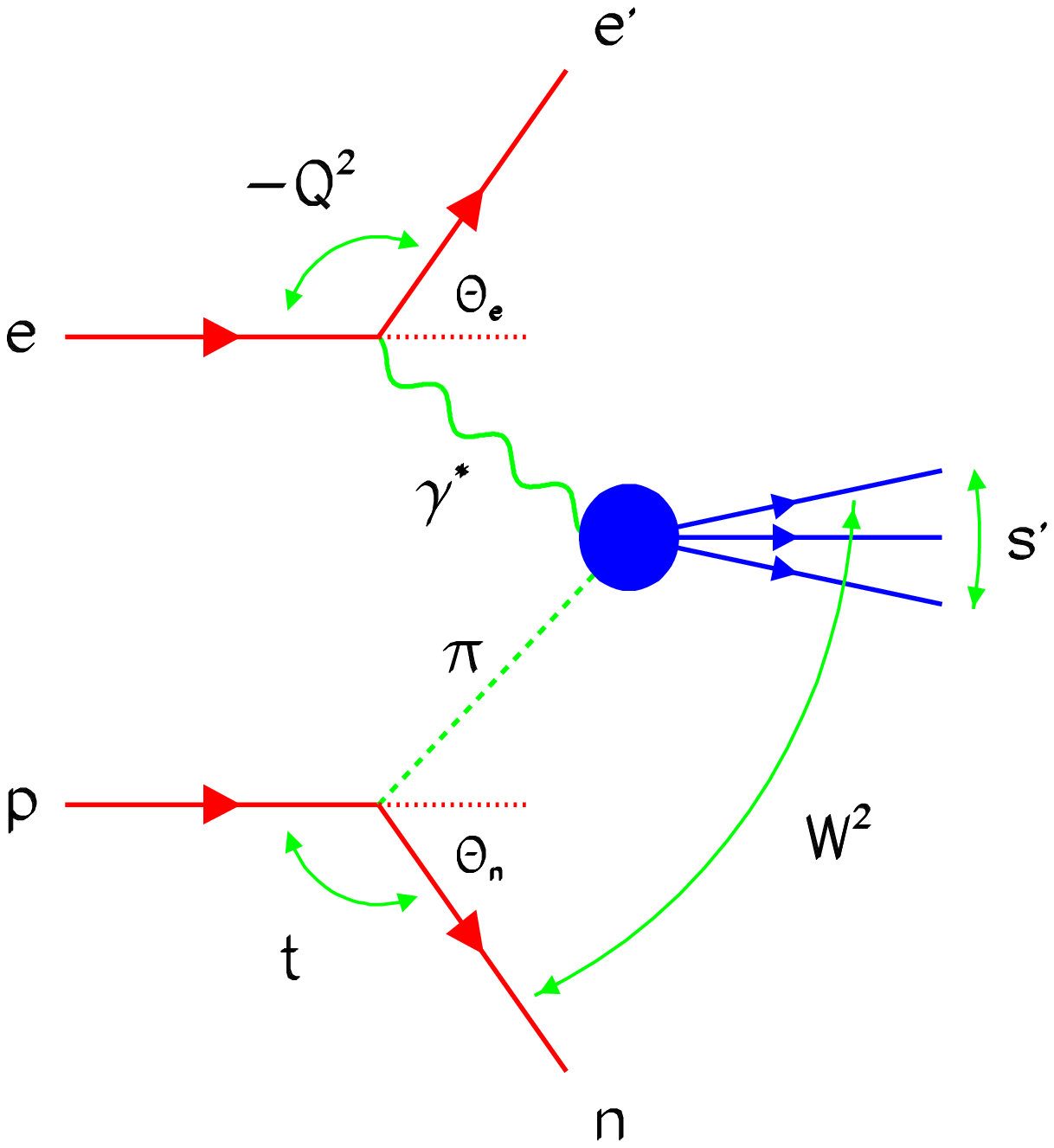,width=.25\textwidth} &
\epsfig{figure=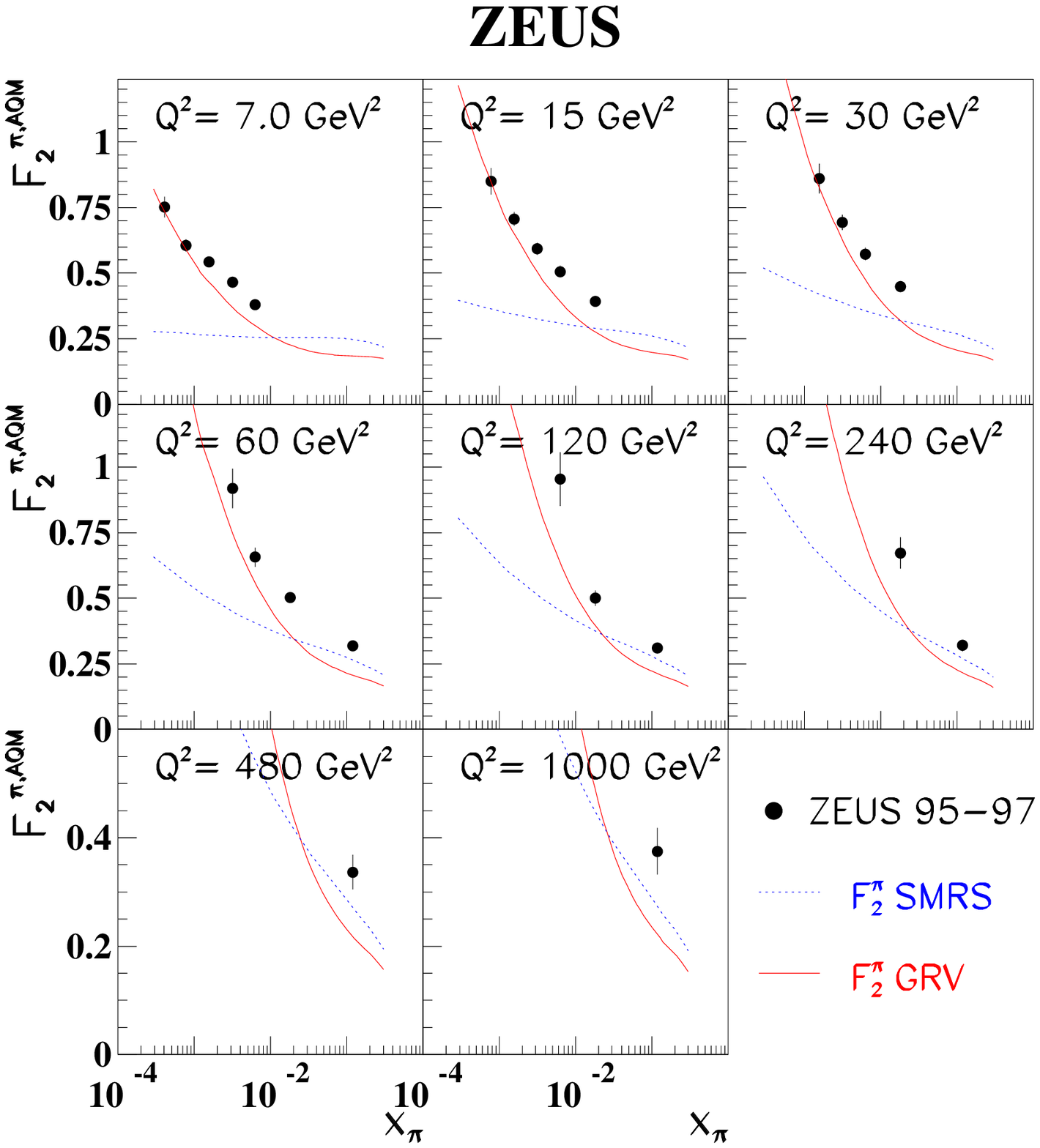,width=.5\textwidth} 
\end{tabular}
\caption{\it $ep\to enX$ at HERA:
(a) diagram for pion exchange; (b) the pion form factor extracted from 
the data at small $t$. From \cite{zeus-epenX}. }
\label{fig:f2pi}
\end{center}
\end{figure}
published data on the process $ep\to enX$ (\cite{zeus-epenX,h1-epenX}). The
neutron emerges at a very small angle with respect to the proton beam direction
and is measured in a dedicated neutron counter at a distance of the order
of $100\,$m downstream from the main detectors. Knowing the energy and angle
of the neutron allows the momentum transfer squared at the $pn$ vertex, $t$, 
to be calculated. At very small values of $t$, the process is dominated by
charged pion exchange as shown in the LH diagram of Fig.~\ref{fig:f2pi}.
In this region of phase space the cross-section for $\gamma^*p\to nX$ may be
written as a convolution of the flux of pions in the proton $f_{\pi/p}$ with
the pion structure function $F_2^\pi$. The largest uncertainty in the 
measurement of $F_2^\pi$ is the pion flux. Current models differ by up to
a factor of two, but this does not affect the $x$ dependence which is shown
in the RH plots of Fig.~\ref{fig:f2pi} - in this case $f_{\pi/p}$
has been estimated using the additive quark model. The data show very
clearly that $F_2^\pi$ rises strongly for $x$ values below 0.01. Thus at
a qualitative level the data support the idea of a universal behaviour at 
small $x$.

\section{Diffraction}
\begin{floatingfigure}[r]{5cm}
\centerline{\epsfig{file=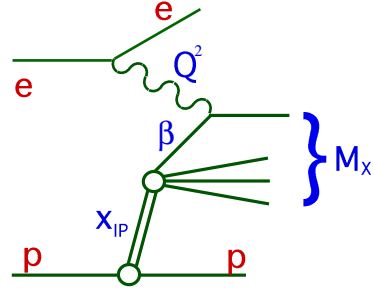,height=3.8cm}}
 \caption{\it 
    Diagram and additional kinematic variables for diffraction
at HERA.
 \vspace{0.5cm}
    \label{fig:dis-diffr} }
\end{floatingfigure}
\noi
The diffractive process $ep\to epX$ is shown in Fig.~\ref{fig:dis-diffr}, 
together with the additional variables needed to describe it beyond $x$,
$Q^2$ and $W^2$ already defined: $x_P\approx 1-p^\prime_z/p_z$, the
fractional longitudinal momentum loss of the proton; $M_X$ the invariant
mass of the diffractive final state at the $\gamma^*$ vertex;
$\beta=Q^2/(2q.(p-p^\prime))$ is the equivalent of Bjorken $x$ in the 
fully inclusive case. Here the focus is on the $W^2$ and $Q^2$
dependence of two classes of diffractive processes: vector meson production 
for which $M_X=M_V$; inclusive diffraction with $M_X>3\,$GeV. Diffractive
events are identified either by direct measurement of the scattered proton
using a `leading proton spectrometer' placed very close to the beam line
at a large distance from the primary interaction or by the presence of a
large rapidity gap in the main detector between the proton direction and
the first energy deposits from the particles making up the $M_X$ system.
As already mentioned, diffraction is a quasi-elastic process and in many
models the underlying physics is closely related to that of the elastic 
process. In particular in the Regge approach, the high energy behaviour of the 
total diffractive cross-sections should be controlled by Pomeron exchange.
\medskip

\subsection{Vector Mesons}

\begin{figure}[tbp]
%\begin{center}
\begin{tabular}{ll}
\epsfig{figure=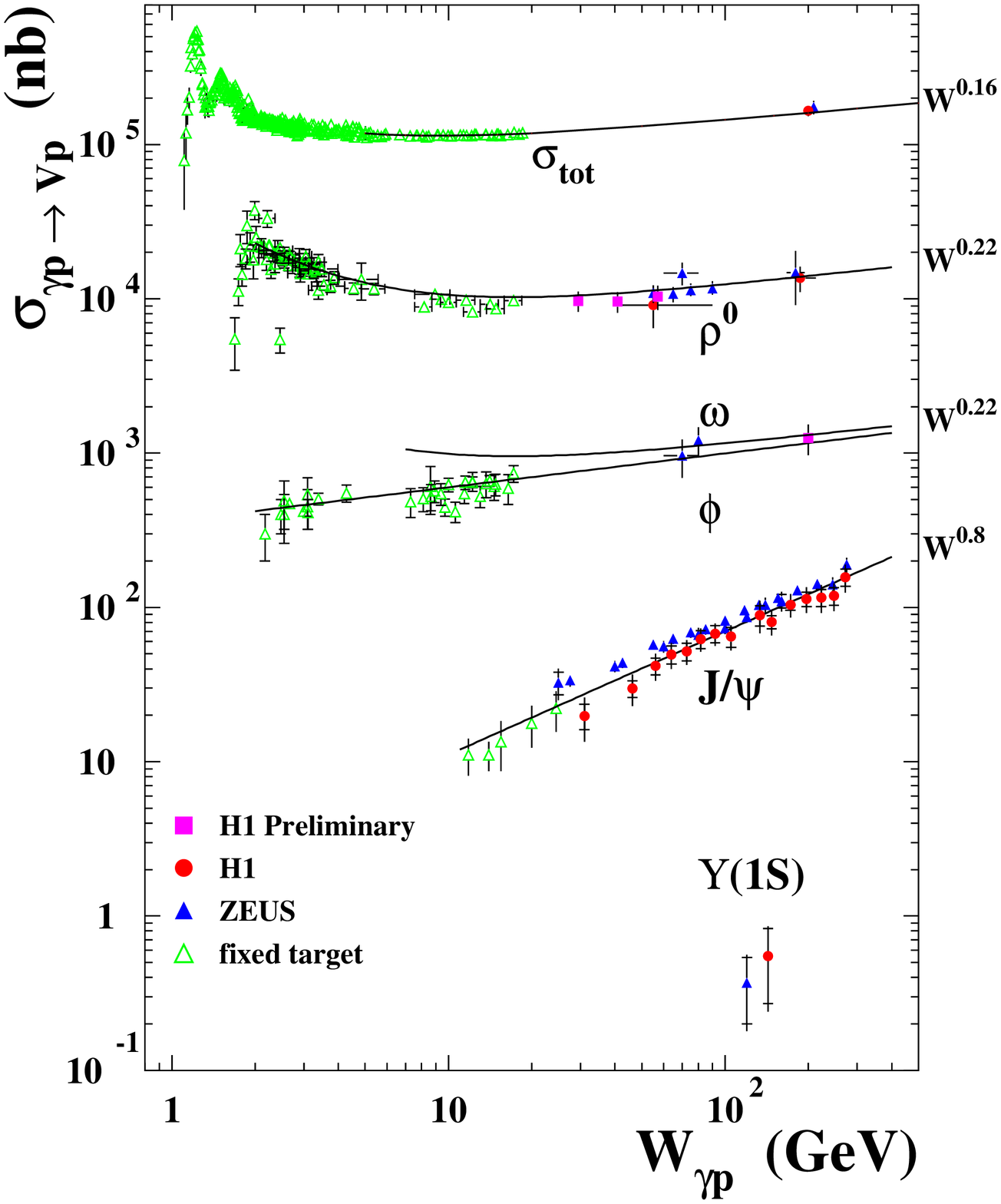,width=.45\textwidth} &
\epsfig{figure=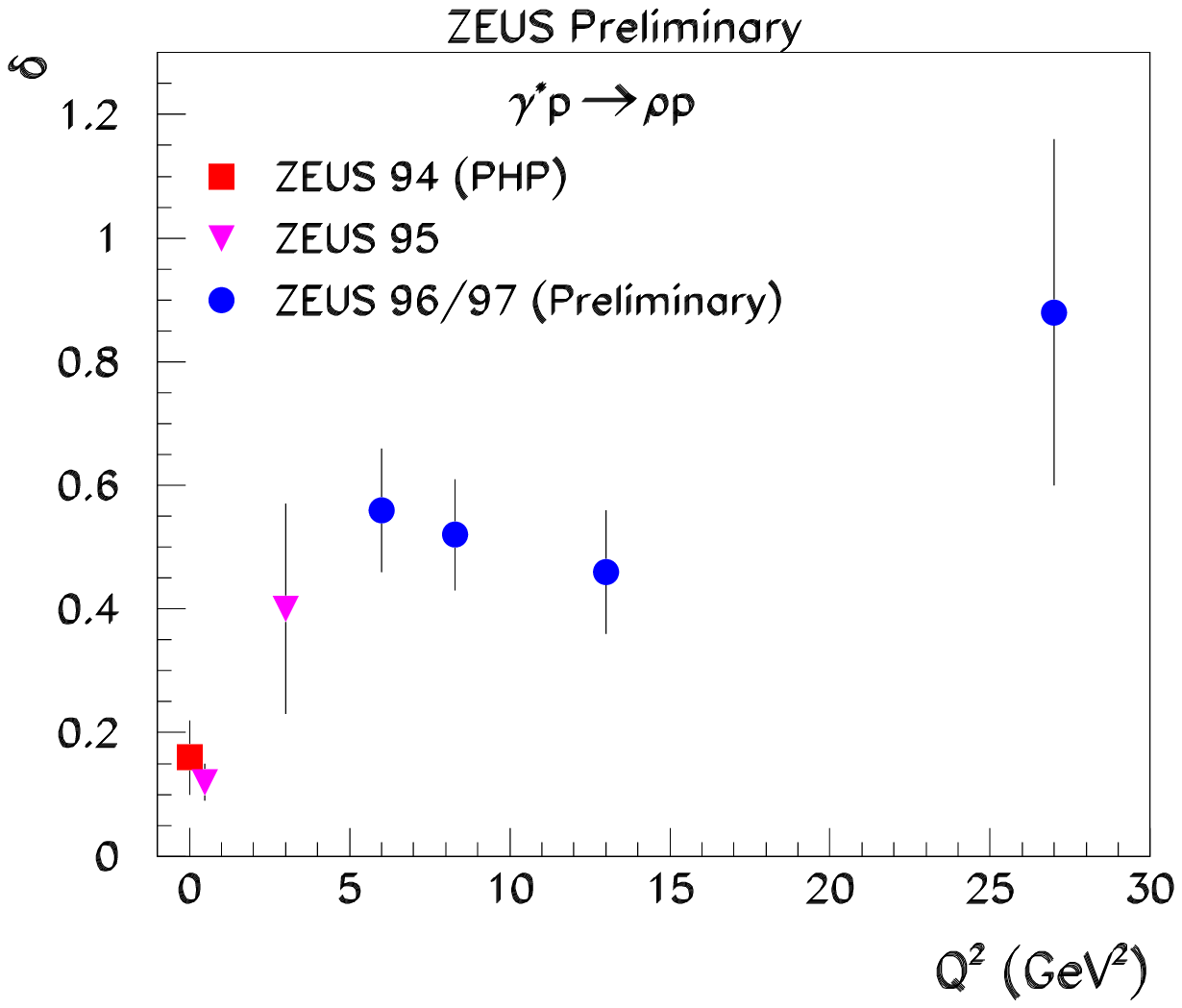,width=.4\textwidth}
\end{tabular}
\caption{\it LH plot, energy dependence of quasi-elastic
vector meson photoproduction; RH plot, exponent of the energy dependence of
$\sigma(\gamma^*p\to \rho^0p)$ vs $Q^2$.
}
\label{fig:diffr-data}
%\end{center}
\end{figure}

First consider the data shown in the LH plot of Fig.~\ref{fig:diffr-data},
$\sigma_{tot}(\gamma p\to Vp)$ for real photoproduction of vector mesons
and $\sigma_{tot}(\gamma p\to X)$ and for $W>10\,$GeV the data are fit to 
a power law dependence $\sigma\sim W^\delta$. For the light vector mesons
($\rho^0,\omega,\phi$) $\delta\sim 0.22$, not inconsistent with the value
expected from Regge theory.
%\footnote{As the diffractive total cross-section 
%is derived from a squared amplitude whereas the fully inclusive total 
%cross-section is given by the imaginary part of the amplitude alone and the 
%energy dependence of diffractive data tends to be plotted against $W$ rather 
%than $W^2$ one might naively expect $\delta\approx 4\lambda$.} 
For $\gamma p\to J/\psi p$ the energy
dependence is much steeper, giving $\delta\sim 0.8$. The RH plot shows
$\delta$ from the energy dependence of $\gamma^*p\to \rho^0p$ in fixed
$Q^2$ bins from 0 to $27\,$GeV$^2$. Although the errors are large, there
is a clear tendency for the energy dependence to steepen as $Q^2$
increases. In both cases the trend is clear, when there is a hard scale
present, either a large $M_V$ or a large $Q^2$, then the cross-section
rises more quickly with energy than expected from soft hadronic physics.

\subsection{Inclusive Diffraction}

Fig.~\ref{fig:diffr+dpol} -- LH  shows a ZEUS measurement of the ratio 
of the diffractive cross-section to the total $\gamma^*p$ cross-section 
for two values of $M_X$ and 8 values of $Q^2$ between $0.27$ and 
$60\,$GeV$^2$ as a function of $W$ \cite{zeus-diffr-ratio}. Assuming 
Pomeron dominance, Regge theory predicts that the energy dependence of the 
ratio is given by $(W^2)^{2\bar{\alpha}_P-\alpha_P(0)-1}$ where 
$\bar{\alpha}_P$
is the value of the Pomeron trajectory evaluated at the mean $t$ of the
diffractive data, for the hadronic Pomeron this gives an
expectation of $W^{0.19}$. Fitting the ratio data at fixed $M_X$ and $Q^2$
to a form $W^\delta$ gives the results
$\delta=0.24\pm 0.07$ for data with $Q^2<1\,$GeV$^2$ and
$\delta=0.00\pm 0.03$ for data with $Q^2>1\,$GeV$^2$. Thus, once again, at
$Q^2$ close to zero the data follows the expectation derived
from soft hadronic physics, whereas in the deep inelastic region this is not
the case and the diffractive cross-section (at fixed $M_X$ and $Q^2$)
follows the energy dependence of $\sigma^{tot}_{\gamma^*p}$.
\medskip

\section{Colour Dipole Models}

Many models have been proposed to describe the transition from the soft
hadronic-like energy dependence at $Q^2\approx 0$ to the harder
behaviour seen at larger $Q^2$. A particularly appealing approach is that
offered by colour dipole models. In the rest frame of the proton
the virtual photon splits into a $q\bar{q}$ pair a long time (or equivalently
a large distance) before the interaction. The $q\bar{q}$ pair, characterised
by a transverse size $r$ and sharing the longitudinal momentum of the proton
in the ratio $(1-z):z$, then interacts with the proton with the 
cross-section $\sigma_{qq}(x,r)$ giving
\be
\sigma^{tot}_{\gamma^*p}(W^2,Q^2)=\int d^2\br dz\,\Psi^*_{\gamma^*}
(\br,z,Q^2)\,\sigma_{qq}(x,r)\,\Psi_{\gamma^*}(\br,z,Q^2),
\ee 
where $\Psi_{\gamma^*}$ is the known wave function for  
$\gamma^*\to q\bar{q}$. The dipole cross-section has to be modelled
\begin{figure}[htbp]
\begin{center}
%\begin{tabular}[t]{ll}
\begin{tabular}{ll}
\epsfig{file=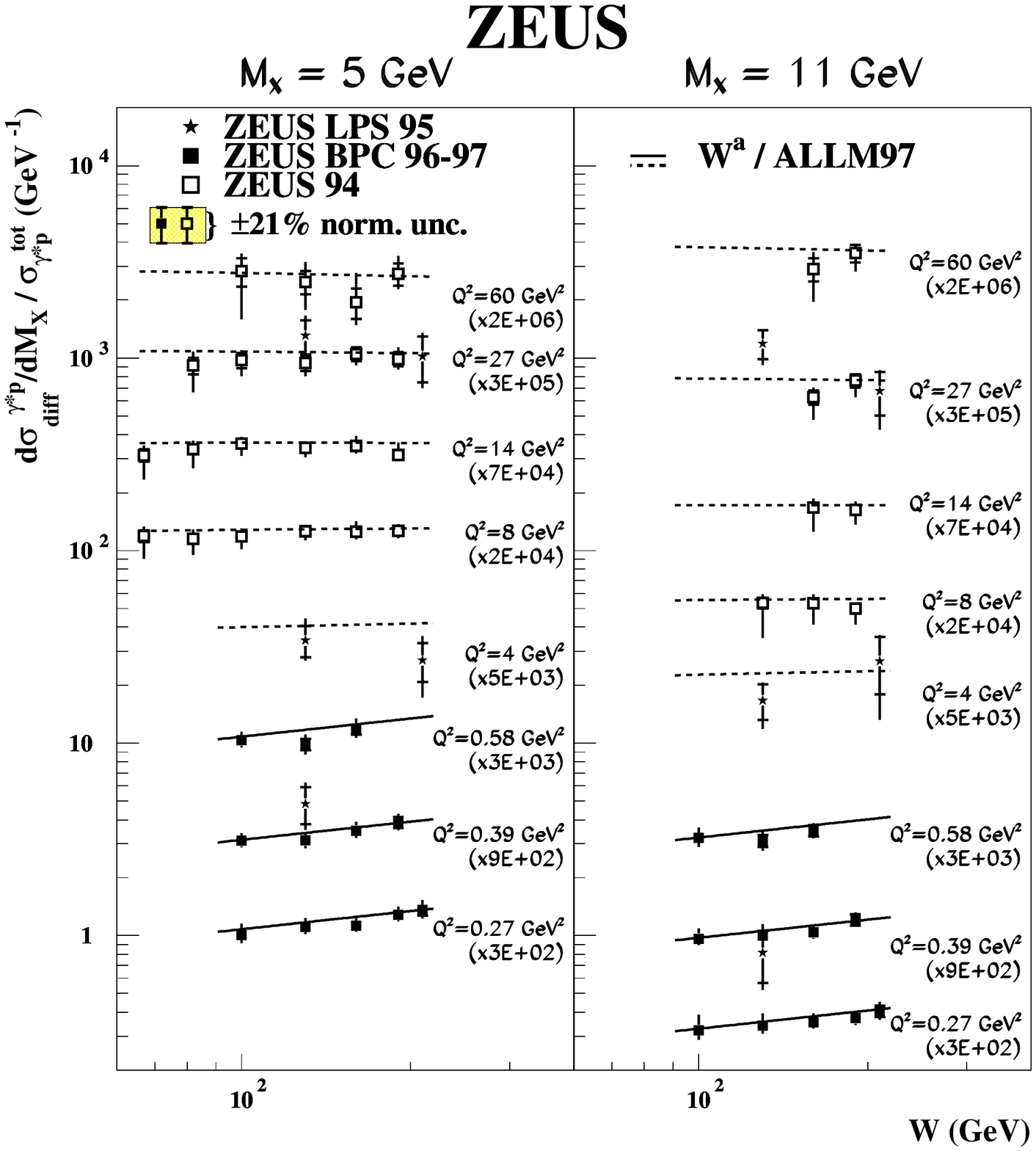,width=.5\textwidth} &
\epsfig{figure=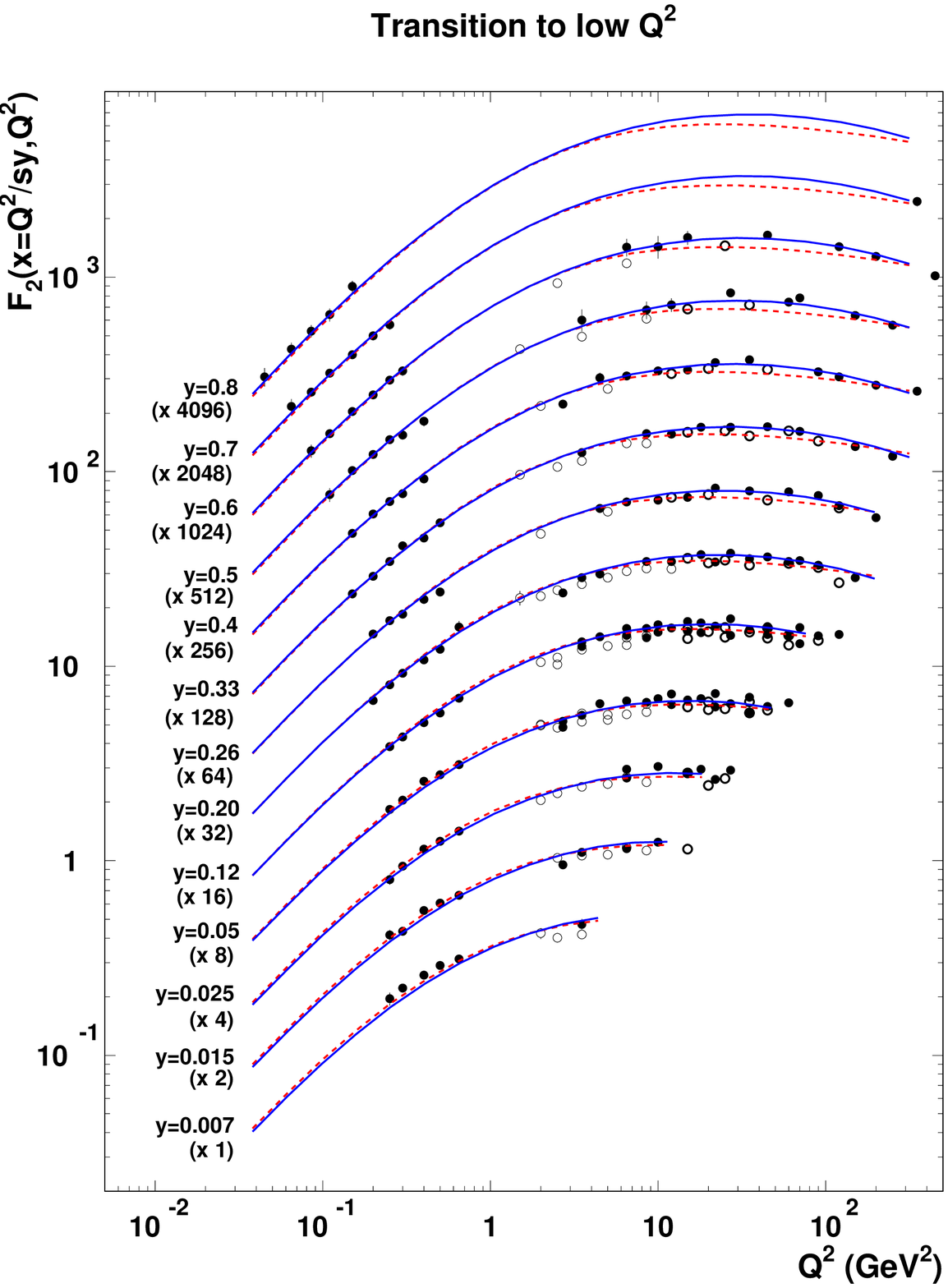,width=.48\textwidth} 
\end{tabular}
\caption{\it LH plots: Ratio of inclusive diffraction to 
$\sigma^{tot}_{\gamma^*p}$ versus $W$ at fixed $M_X$ and $Q^2$.
RH plot: The colour dipole model of Golec-Biernat \etal fit to $F_2$ data
in bins of constant $y$.}
\label{fig:diffr+dpol}
\end{center}
\end{figure}
and a number of formulations have been proposed (see \cite{forshaw}
and references therein).
Here the version of Golec-Biernat \& W\"usthoff (GBW) \cite{gbw1} is followed,
partly for its simplicity, but also because it builds in the constraint
of saturation. In their approach,
the dipole cross-section is given by 
\be
\sigma_{qq}(x,r)=\sigma_0\left[1-\exp(-r^2/4R_0^2(x))\right],~~~
R_0(x)=(x/x_0)^{\lambda/2},
\label{eqn:gbw1}
\ee
where $\sigma_0, x_0, \lambda$ are three parameters to be determined from 
data. The interesting features of Eq.~\ref{eqn:gbw1} are: for small
dipole sizes $r<<R_0,~~\sigma_{qq}\propto r^2x^{-\lambda}$; for
large dipoles $r>>R_0, ~~\sigma_{qq}\propto \sigma_0$ and that the parameter
$R_0$ setting the scale depends on $x$. This last feature means that 
the approach to the `saturation limit' occurs for smaller dipole sizes
as $x$ decreases. The parameters of the model are fixed by fitting the
HERA low $x$ $F_2$ data. Another very compelling feature of the dipole
approach is that diffraction may also be described. The diffractive
cross-section is given by
\be
\left.{d\sigma^{Diffr.}_{\gamma^*p}\over dt}\right|_{t=0}=
{1\over 16\pi}\int d^2\br dz|\Psi(\br,z)|^2\sigma^2_{qq}(x,r).
\ee
Thus no new parameters are required beyond those needed to describe the
total $\gamma^*p$ cross-section. The successes of this simple model are 
a reasonable fit to the low $x$ and low $Q^2$ (including the transition
to the non pQCD region - see Fig.~\ref{fig:diffr+dpol}-RH) 
and the energy independence of the
ratio $\sigma^{Diffr.}_{\gamma^*p}/\sigma^{tot}_{\gamma^*p}$ \cite{gbw2}. 
An obvious deficiency of the model is that it does not contain any
mechanism for pQCD $Q^2$ evolution of the structure function. An attempt
to remedy this has been made by Bartels \etal \cite{gbbk}, the dipole
cross-section for small $r$ is modified and related to the gluon density
by
\be
\sigma_{qq}(r,x)\approx {\pi^2\over 3}r^2\alpha_S\,xg(x,\mu^2),
\ee
where the scale $\mu^2\approx C/r^2$ and $C$ is a parameter.
The modified model now has 5 parameters to be determined by the $F_2$
data. The fit to the higher $Q^2$ data is improved and $\lambda(Q^2)$ 
(Eq.~\ref{eqn:lam-def}) is well described for all $Q^2$, whereas for 
the non-evolving model $\lambda$ falls below the slope data for 
$Q^2>10\,$GeV$^2$. The success of  the GBW dipole model cannot be 
taken to imply that
saturation is {\it required} by the HERA data as other models without
saturation, e.g. the dipole model of Forshaw \etal \cite{forshaw}, give
as good a representation of the data.

\section{Summary}

HERA has provided high precision data on the proton structure function
$F_2$ over a wide range of $x$ and $Q^2$. In the low $x$ region the
bulk of the data is now systematics limit (typical uncertainties 
$\sim 2$\%). HERA has also provided a wide range of measurements on
hard diffractive scattering and quasi-elastic vector meson production.
The striking feature of $F_2$ at low $x$ -- the strong rise -- is 
mirrored in diffractive processes when there is a hard scale. The first
measurement of the pion structure function at low $x$ shows that it
too rises strongly as $x$ decreases, which hints at universality in
low $x$ dynamics. The behaviour of low $x$ $F_2$, or equivalently high energy 
$\sigma^{tot}_{\gamma^*p}$, is dominated by gluon dynamics. Although
some of the important calculations were completed well before HERA 
started operations, there is no doubt that HERA has opened up new
avenues in strong interaction physics. Particularly the determination
of the gluon density at low $x$, the refinement of high density 
perturbative gluon dynamics and the deepening of the 
relationship between diffractive scattering and the physics underlying
the rise of total cross-sections with energy. The HERA measurements
and the related theoretical developments provide essential input for
Run II at the Tevatron, RHIC\footnote{See the contribution by Steinberg
\cite{steinberg}.} and the LHC. 

\section{Acknowledgements}

Thanks to my colleagues in ZEUS, H1 and the HERA `low $x$ club'
for providing real and virtual help in the preparation of this talk.

\end{document}